\documentclass[prb,aps,eqsecnum,showpacs,twocolumn]{revtex4}

\usepackage[dvips]{graphics}
\usepackage{graphicx}
\usepackage{amssymb}
\usepackage{epstopdf}
\usepackage{bm}

\begin{document}


\title{Odd triplet superconductivity in clean and moderately disordered SFFS junctions}
\author{Zorica Popovi\'c and Zoran Radovi\'c}
\address{Department of Physics, University of Belgrade, P.O. Box 368, 11001 Belgrade, Serbia}

\begin{abstract}
We study the Josephson effect and pairing correlations in SFFS junctions that
consist of  conventional superconductors (S) connected through two
metallic monodomain ferromagnets (F) with transparent and spin inactive
interfaces. We solve the Eilenberger equations for arbitrary
relative orientation of  magnetizations of the two F layers in the clean limit and for moderate disorder in ferromagnets. Spatial variation of pair amplitudes, singlet $f_s$, and odd in frequency triplet  $f_{t0}$ and $f_{t1}$, with 0 and $\pm1$ spin projections, as well as the Josephson current-phase relations are calculated for different values of the ferromagnetic layers thickness and angle $\alpha$ between in-plane  magnetizations.  In contrast to the dirty limit case, we find that for $0<\alpha<\pi$ both spin singlet and triplet pair
amplitudes in F layers power-law decay in the same oscillatory manner with distance from
the FS interfaces. This decay gets faster as the impurity-scattering rate in ferromagnets is increased. The computed triplet amplitude $f_{t1}$ has opposite signs in the two magnet regions, penetrates into the superconductors and monotonically decays over the same distance, which is the superconducting coherence length, as the singlet amplitude $f_s$ saturates to the bulk value. We point out that influence of misorientation of magnetizations  on the Josephson current can not be attributed directly to the appearance of odd triplet correlations.

\end{abstract}


\maketitle

\section{Introduction}
Odd-frequency triplet superconducting correlations with non-zero total spin projection induced by proximity effect in
heterostructures containing superconductors with ordinary singlet pairing and inhomogeneous ferromagnets  have been of considerable interest in the last decade.~\cite{bergerettriplet,bergeret03prltriplet,reviewbergeret,Keizer,Eschrig08,asano} Odd-frequency pairing mechanism was proposed long ago in an attempt to describe the A phase of superfluid $^3\rm{He}$.~\cite{berezinski} Although this superconducting state is thermodynamically stable,~\cite{Martin} it was found that in the case of superfluid $^3\rm{He}$ the pairing is odd in space ($\it{p}$-wave) rather than in time, "odd" triplet superconductivity occurs in certain superconductor-ferromagnet (SF) structures.~\cite{Keizer} It is believed that this exotic $\it{s}$-wave pairing state which is even in momentum, but with the triplet correlations being odd in frequency, can enhance significantly the superconducting penetration length at the SF interface,~\cite{reviewbergeret} providing a Josephson current through half-metallic barriers.~\cite{Keizer,Eschrig08,asano} The simplest examples of SF heterostructures with inhomogeneous magnetization are FSF and SFFS heterojunctions with homogeneous monodomain ferromagnetic layers having an angle $\alpha$ between their in-plane magnetizations. These structures have been studied using quasiclassical approach in both diffusive~\cite{bergerettriplet,bergeret03prltriplet,reviewbergeret,Golubov,You,lof,barash02,sudbo08} and clean~\cite{Eschrig03,blanter04,sudbo09} limits by solving Usadel~\cite{usadel71} and Eilenberger~\cite{eilenberger68} equations, respectively, and by solving the Bogoliubov-de Gennes equation.~\cite{Pajovic,Milos,asano,Valls07,Valls08}
In the case of parallel ($\alpha=0$) magnetizations, SFS junctions,~\cite{radovic03,Buzdin08} besides spin-singlet superconducting correlations, only odd triplet correlations with zero total spin projection exist. These correlations penetrate into the ferromagnet over a short length scale determined by the exchange energy.  For noncollinear magnetizations, triplet correlations with nonzero spin projection are present as well. It is expected  that they are not suppressed by the exchange interaction, and consequently that they are long ranged.~\cite{bergeret03prltriplet} It has been predicted for diffusive junctions that long-range spin-triplet components with spin projection $\pm 1$ should have a dramatic impact on transport properties and the Josephson effect, displayed through a nonmonotonic dependence on angle between magnetizations. In diffusive Josephson junctions, the length scales associated with short- and long-range correlations are, respectively, $\xi_F^d=\sqrt{\hbar D_F/h_0}$ and $\xi_N^d=\sqrt{\hbar D_F/k_BT}$, where $D_F$ is the diffusion constant in the ferromagnet, and thermal energy $k_BT$ is typically much smaller than the exchange energy $h_0$. However, this is not the case in ballistic SF heterostructures  in which the only characteristic length is the ferromagnet coherence length $\xi_F=\hbar v_F/h_0$, where $v_F$ is the Fermi velocity. In the clean limit, influence of misorientation of magnetizations  on the Josephson current or conductance can not be attributed to the triplet correlations.~\cite{Pajovic,Milos} Moreover, for noncollinear magnetizations, $\alpha\neq 0$ and $\pi$, both spin singlet and triplet pair amplitudes in  ferromagnetic layers decay in the same oscillatory manner with distance from the FS interfaces.~\cite{Valls07,Valls08}

In this paper, we study the Josephson effect and odd-frequency triplet superconductivity in clean and moderately disordered SFFS junctions where the magnetic interlayer consists of two monodomain ferromagnets having  relative angle $\alpha$ between their in-plane magnetizations. Solving the Eilenberger equations we calculate singlet and triplet pair amplitudes, and the Josephson current. In contrast to the dirty limit case we find that for $0<\alpha<\pi$, both spin singlet and triplet pair
amplitudes in F layers are oscillating and power low decaying with distance from
the FS interfaces. This decay gets faster as the impurity-scattering rate in ferromagnets is increased. The computed triplet amplitudes have opposite signs in the two magnet regions, penetrate into the superconductors and monotonically decay over the  superconducting coherence length, $\xi_S=\hbar v_F/\pi\Delta_0$, where $\Delta_0$ is the
superconducting pair potential, which is the same distance on which the singlet amplitude saturates to the bulk value. As in the previous work,~\cite{Pajovic} no substantial impact of spin-triplet superconducting correlation on the Josephson current has been found. The critical  value of the Josephson current $I_c$ is a monotonic function of angle $\alpha$ when the junction is far enough from the $0-\pi$ transitions. Otherwise, $I_c(\alpha)$ is a nonmonotonic function of $\alpha$ with characteristic dips related to the onset of $0-\pi$ transitions, but this nonmonotonicity is a consequence of decreasing influence of the exchange potential with increasing misorientation of magnetizations.~\cite{Pajovic} Influence of misorientation of magnetizations  on the Josephson current can not be attributed to the appearance of triplet correlations: with increasing misorientation of magnetizations, $\alpha\rightarrow \pi$, the Josephson currents are the same as for parallel magnetizations, $\alpha=0$, with correspondingly decreasing exchange potential, $h_0\rightarrow 0$. Therefore, rotation of magnetizations in SFFS junctions is equivalent to the decreasing influence of the homogeneous exchange potential, in the absence of triplet correlations.

The paper is organized as follows. In Sec. II we present the model and the solutions of Eilenberger equations that we used to calculate the Josephson current and spin-singlet and -triplet pair amplitudes. In Sec. III we present the numerical results, and the conclusion in Sec. IV.

\section{Model and formalism}
\begin{figure}
\includegraphics[width= 7cm]{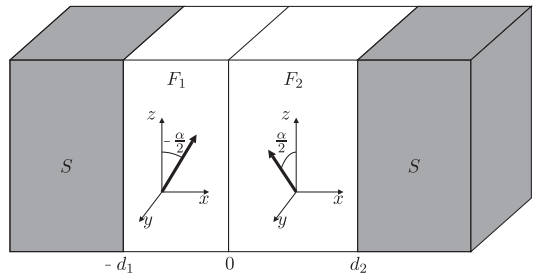}
\caption{Schematics of an SF$_1$F$_2$S heterojunction. The
magnetization vectors lie in the $y$-$z$ plane and form the opposite
angles $\pm \alpha/2$ with respect to the $z$-axis.}
\label{Fig1}
\end{figure}

We consider a clean SF$_1$F$_2$S  heterojunction consisting of two
superconductors (S), and two uniform mono-domain ferromagnetic
layers (F$_1$ and F$_2$) of thickness $d_1$ and $d_2$, with
misorientation angle $\alpha$ between their in-plane magnetizations (see
Fig.~\ref{Fig1}). Interfaces between layers are fully transparent
and magnetically inactive. Superconductors are described in the
framework of quasiclassical theory of superconductivity, while the
ferromagnetism is modeled by the Stoner model, using an exchange
energy shift $2h_0$ between the spin subbands. Disorder is
characterized by the mean free path $l=v_F\tau$, where $\tau$ is the
average time between  scattering on impurities and $v_F$ is the
Fermi velocity assumed to be the same everywhere. We consider the
clean limit and moderately diffusive case when $l$ is larger than the two characteristic lengths:
the ferromagnetic exchange length $\xi_F=\hbar v_F/h_0$, and the superconducting
coherence length $\xi_S=\hbar v_F/\pi\Delta_0$, where $\Delta_0$ is the
superconducting pair potential.

In this model the Eilenberger Green functions
$g_{\sigma\sigma^{'}}(x,v_x,\omega_n)$, $f_{\sigma\sigma^{'}}(x,v_x,\omega_n)$, and
$f^{+}_{\sigma\sigma^{'}}(x,v_x,\omega_n)$  depend on the center-of-mass
coordinate $x$ along the junction axis $x$, on the angle $\theta$
of the quasiclassical trajectories with respect to  $x$ axis,
$v_x=v_F\cos{\theta}$ being the projection of the Fermi velocity
vector, and on the Matsubara frequencies $\omega_n=\pi k_BT(2n+1)$, $n=0,\pm1,\pm2,$ etc. Spin indexes are $\sigma=\uparrow,\downarrow$.

The Eilenberger equation in spin$\otimes$particle-hole space can be
written in the compact form
\begin{equation}
\label{Eil} \hbar v_x\partial_x \check{g}+\Big[\omega_n
\hat{\tau}_3\otimes \hat{1}-i \check{V}+\check{\Delta}+\hbar \langle
\check{g}\rangle/2\tau, \check{g}\Big]=0,
\end{equation}
with normalization condition $\check{g}^2=\check{1}$. We indicate by $\hat{\cdots}$ and $\check{\cdots}$ $2\times2$ and $4\times4$ matrices, respectively. The brackets
$\langle\ldots\rangle$ denote angular averaging over the Fermi surface (integration over $\theta$) and
$[\ldots]$ denotes a commutator. Here, the quasiclassical Green
functions
\begin{equation}
\check{g}=\left[
\begin{array}{rrrr}
g_{\uparrow\uparrow} & g_{\uparrow\downarrow} & f_{\uparrow\uparrow} & f_{\uparrow\downarrow}\\
g_{\downarrow\uparrow} & g_{\downarrow\downarrow} &
-f_{\downarrow\uparrow} & f_{\downarrow\downarrow}\\
f_{\uparrow\uparrow}^{+} & f_{\uparrow\downarrow}^{+} &
-g_{\downarrow\downarrow} & -g_{\downarrow\uparrow}\\
-f_{\downarrow\uparrow}^{+} & f_{\downarrow\downarrow}^{+} &
-g_{\uparrow\downarrow} & -g_{\uparrow\uparrow}
\end{array}  \right],
\end{equation}
are related to the corresponding Gor'kov-Nambu Green functions $\check{G}$ integrated
over the energy $\varepsilon_{\mathbf{k}}=\hbar^2k^2/2m-\mu$,
\begin{equation}
\check{g}=\frac{i}{\pi}\int
d\varepsilon_\mathbf{k}\vspace{2mm}\check{G}.
\end{equation}
The matrix $\check{V}$ is given by

\begin{eqnarray}
\check{V}=\hat{1}\otimes \textrm{Re}\Big[
\textbf{h}(x)\cdot\widehat{\bm{\sigma}}\Big]+i\hat{\tau}_3\otimes
\textrm{Im}\Big[ \textbf{h}(x)\cdot\widehat{\bm{\sigma}}\Big],
\end{eqnarray}
 where components  $\hat{\sigma}_x,
\hat{\sigma}_y, \hat{\sigma}_z$ of the vector
$\widehat{\bm{\sigma}}$ and $\hat{\tau}_1, \hat{\tau}_2,
\hat{\tau}_3$ are Pauli matrices in spin and particle-hole space,
respectively. The
in-plane, $y$-$z$, magnetizations of the neighboring F layers are
not collinear in general, and the magnetic domain structure is
described by the angle $\mp{\alpha}/2$ with respect to the $z$-axis,
for the left (F$_1$) and right(F$_2$) ferromagnets, respectively.
For simplicity, we assume equal magnitude of the exchange
interaction in ferromagnetic domains,
$\textbf{h}(x)=h_0(0,\mp\sin{\alpha/2},\cos{\alpha/2})$, and
$\Delta(x)=0$ for $-d_1<x<d_2$.

The matrix $\check{\Delta}$ is
\begin{equation}
\check{\Delta}=\left[
\begin{array}{cc}
0 & \hat{\tau}_1\Delta\\
\hat{\tau}_1\Delta^{*} & 0
\end{array}  \right].
\end{equation}
We assume that the superconductors are identical and, in the step-wise (non self-consistent) approximation, we take the pair potential $\Delta$
in the form
\begin{equation}
\Delta=\Delta_0 \left[ e^{-i\phi /2}\Theta (-x-d_{1})+e^{i\phi
/2}\Theta (x-d_{2})\right] ,
\end{equation}
where $\Delta_0 $ is the bulk superconducting gap and $\phi$ is the
macroscopic phase difference across the junction. The temperature
dependence of $\Delta_0 $ is given by $\Delta_0 (T)=\Delta_0
(0)\tanh \left( 1.74\sqrt{T_{c}/T-1}\right)$.\cite{munhl} Although the
self-consistent calculations are needed when the proximity effect is
strong between S and F layers,~\cite{Valls07} one would not
expect this to qualitatively change the spatial variation of the
Green functions.

In the clean limit, $\tau\rightarrow \infty$, solutions of Eq.~(\ref
{Eil}) for the left superconductor ($x<-d_1$) can be written in the
usual form for normal Green functions
\begin{eqnarray}
\label{gS}
g_{\uparrow\uparrow}(x)&=&\frac{\omega_n}{\Omega_n}+D_1e^{\kappa_sx},\\
g_{\uparrow\downarrow}(x)&=&D_2e^{\kappa_sx},\\
g_{\downarrow\uparrow}(x)&=&D_3e^{\kappa_sx},\\
g_{\downarrow\downarrow}(x)&=&\frac{\omega_n}{\Omega_n}+D_4e^{\kappa_sx},
\end{eqnarray}
 and for anomalous Green functions
\begin{eqnarray}
f_{\uparrow\uparrow}(x)&=&\frac{2\Delta}{2\omega_n+\hbar v_x\kappa_s}D_2e^{\kappa_sx},\\
f_{\uparrow\downarrow}(x)&=&\frac{\Delta}{\Omega_n}+\frac{2\Delta}{2\omega_n+\hbar v_x\kappa_s}D_1
e^{\kappa_sx},\\
f_{\downarrow\uparrow}(x)&=&-\frac{\Delta}{\Omega_n}-
\frac{2\Delta}{2\omega_n+\hbar v_x\kappa_s}D_4e^{\kappa_sx},\\
f_{\downarrow\downarrow}(x)&=&\frac{2\Delta}{2\omega_n+\hbar v_x\kappa_s}D_3e^{\kappa_sx},\label{fS}
\end{eqnarray}
 with
\begin{equation}
\kappa_s=\frac{2\Omega_n}{\hbar v_F|\cos(\theta)|},
\end{equation}
and $\Omega_n=\sqrt{\omega_n^2+|\Delta|^2}$. For the right
superconductor ($x>d_2$), the solutions keep the same form with
$\kappa_s\longrightarrow -\kappa_s$ and new set of constants
$D_1^{'},...,D_4^{'}$.

Solutions for the Green functions in the clean limit,
$\tau\rightarrow \infty$, for the left ferromagnetic layer F${_1}$,
$-d_1<x<0$, can be written in the form
\begin{widetext}
\begin{eqnarray}
\label{gF}
g_{\uparrow\uparrow}(x)&=&K_1+i\tan(\alpha/4)K_2e^{i\kappa_{0}x}- i \tan(\alpha/4)K_3e^{-i\kappa_{0}x},\\
g_{\uparrow\downarrow}(x)&=&K_2e^{i\kappa_{0}x}+
\tan^2(\alpha/4)K_3e^{-i\kappa_{0}x}+\frac{i}{2}\tan(\alpha/2)[K_1-K_4],\\
g_{\downarrow\uparrow}(x)&=&K_3e^{-i\kappa_{0}x}+
\tan^2(\alpha/4)K_2e^{i\kappa_{0}x}-\frac{i}{2}\tan(\alpha/2)[K_1-K_4],\\
g_{\downarrow\downarrow}(x)&=&K_4-
i\tan(\alpha/4)K_2e^{i\kappa_{0}x}+
i\tan(\alpha/4)K_3e^{-i\kappa_{0}x},\label{gF11}
\end{eqnarray}
and
\begin{eqnarray}
\label{f1}
f_{\uparrow\uparrow}(x)&=&C_1e^{-\kappa x}- i\tan(\alpha/4)C_2e^{-\kappa_{-}x}-i \tan(\alpha/4)C_3e^{-\kappa_{+}x},\\
f_{\uparrow\downarrow}(x)&=&C_2e^{-\kappa_{-}x}-\tan^2(\alpha/4)C_3e^{-\kappa_{+}x}-
\frac{i}{2}\tan(\alpha/2)
[C_1+C_4]e^{-\kappa x},\\
f_{\downarrow\uparrow}(x)&=&-C_3e^{-\kappa_{+}x}+\tan^2(\alpha/4)C_2e^{-\kappa_{-}x}+\frac{i}{2}\tan(\alpha/2)
[C_1+C_4]e^{-\kappa x},\\
f_{\downarrow\downarrow}(x)&=&C_4e^{-\kappa x}-
i\tan(\alpha/4)C_2e^{-\kappa_{-}x}-
i\tan(\alpha/4)C_3e^{-\kappa_{+}x},\label{fF}
\end{eqnarray}
\end{widetext}
where
 \begin{equation}
\kappa_0=2h_0/\hbar v_x,
\end{equation}
and
\begin{eqnarray}
\label{k1}
\kappa&=&2\omega_n/\hbar v_x,\\
\kappa_{-}&=&2(\omega_n-ih_0)/\hbar v_x,\\
\kappa_{+}&=&2(\omega_n+ih_0)/\hbar v_x.\label{k3}
\end{eqnarray}
The solution for the right ferromagnetic layer F${_2}$, $0<x<d_2$, can
be obtained by substitution $\alpha\longrightarrow -\alpha$, with a
new set of constants $K_1^{'},...,K_4^{'}$ and
$C_1^{'},...,C_4^{'}$. The complete solution requires one to determine $24$ unknown
coefficients: $4+4$ in superconducting electrodes, and
$8+8$ in ferromagnetic layers. Boundary conditions of continuity of $\check{g}$ at the three interfaces provide
necessary 24 equations.

For $h_0=0$ solutions (\ref{gF})-(\ref{gF11}) reduce to the solutions for SNS junctions with $K_2=K_3=0$ and $K_4=K_1$. Solutions (\ref{f1})-(\ref{fF}) reduce to the SNS case if we take $\alpha=0$ and than $h_0=0$ with $C_1=C_4=0$ and $C_3=C_2$. Note that the same solutions are obtained for $\alpha=\pi$ and $h_0\neq 0$.

The supercurrent density is given by the normal Green function through the following expression
\begin{equation}
\label{struja} I(\phi)=\pi e
N(0)k_BT\sum_{\omega_n}\sum_{\sigma}\langle v_x
\textrm{Im}\hspace{1mm} g_{\sigma\sigma}(v_x)\rangle ,
\end{equation}
where $N(0)$ is the density of states per spin at the Fermi level.

The zero-time pair amplitudes,
singlet $f_s$, and triplet $f_{t0}$ and $f_{t1}$, with  0 and $\pm1$ projections of the total spin of a pair, are defined in terms of anomalous Green
functions as
\begin{eqnarray}
f_s&=&\pi N(0)
k_BT\sum_{\omega_n}(f_{\uparrow\downarrow}-f_{\downarrow\uparrow}),\\
f_{t0}&=& \pi N(0)
k_BT\sum_{\omega_n}(f_{\uparrow\downarrow}+f_{\downarrow\uparrow}),\\
f_{t1}&=&\pi N(0)
k_BT\sum_{\omega_n}(f_{\uparrow\uparrow}+f_{\downarrow\downarrow}).
\end{eqnarray}
The ground state in SFS junctions can be $0$ or $\pi$ state.\cite{reviewsSFS} For $\phi=0$ the singlet amplitude $f_s$ is real, while the triplet
amplitudes $f_{t0}$ and $f_{t1}$ are imaginary, and for $\phi=\pi$  the opposite is true. For $0<\phi<\pi$, all the amplitudes are complex functions. In the figures, we will normalize
all amplitudes to the value of $f_s$ in bulk superconductors,
\begin{equation}
f_{sb}=\pi N(0) k_BT\sum_{\omega_n}\frac{\Delta_0}{\Omega_n}.
\end{equation}

The influence of moderate disorder in ferromagnetic layers on the Josephson current and on the
pair amplitudes is studied in the limit $h_0\tau/\hbar\gg 1$. From the term
$\hbar\langle\check{g}\rangle/2\tau$ in Eq.~(\ref{Eil}) we kept the
largest contribution, which is $\langle g_{\sigma\sigma}\rangle$ calculated for $\tau\rightarrow \infty$. The contributions of $\langle f_{\sigma\sigma^{'}}\rangle$ and $\langle g_{\sigma\bar{\sigma}}\rangle$  are neglected.  Introducing  $\langle g_{\uparrow\uparrow}\rangle=R+iI$, we find that $\langle
g_{\downarrow\downarrow}\rangle=R-iI$. Numerical calculations show that
$R$ is independent of $x$, and
$R\approx$ sign$(\omega_n)$ for any $\alpha$. The imaginary part $I$
oscillates with $x$, except for $\alpha=0$, but it is
nontrivial only in the vicinity of $\omega_n=0$. Therefore, the contribution  of $I$ to
the pair amplitudes, as well as to the supercurrent, can be neglected, and the previous solutions Eqs.(\ref{gF})--(\ref{fF}) are used by replacing only $\kappa,\kappa_{\pm}$, Eqs.(\ref{k1})--(\ref{k3}), with

\begin{eqnarray}
\label{k11}
\tilde{\kappa}&=&\kappa+  \frac{R}{\tau v_x},\\
\tilde{\kappa}_{-}&=&\kappa_{-}+  \frac{R}{\tau v_x},\\
\tilde{\kappa}_{+}&=&\kappa_{+}+  \frac{R}{\tau v_x}.\label{k13}
\end{eqnarray}


\section{Results}
For simplicity, we consider a single transfer channel (one-dimensional case). We illustrate our results for symmetric junctions
with relatively thin and thick ferromagnetic layers,
$d_{1}=d_{2}=50k_F^{-1}$ and $500k_F^{-1}$, and for low  temperature $T/T_c=0.1$. Superconductors are
characterized with the zero temperature value of the bulk pair potential  $\Delta_0(0)
/E_{\rm F}=10^{-3}$, which corresponds to $\xi_S(0)=636k_F^{-1}$. In order to achieve quantitatively reliable
results within the quasiclassical approximation, we assume that all interfaces are fully
transparent, the Fermi wave vectors in all metals are equal ($k_F$), and the ferromagnets
are relatively weak, $h_0/E_F=0.1$, which corresponds to $\xi_F=20k_F^{-1}$.

\begin{figure}
\includegraphics[width= 7cm]{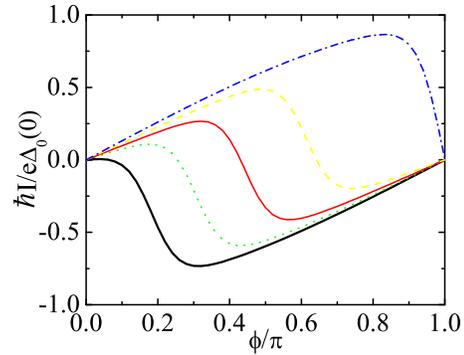}
\caption{(Color online) The current-phase relation $I(\phi)$ in the clean limit for
$T/T_c=0.1$, $h_0/E_{\rm F}=0.1$, $d_1=d_2=50k_{\rm F}^{-1}$,
and five  values of the misorientation angle: $\alpha=0$ (thick solid
curve), $\pi/3$ (dotted curve),  $\pi/2$ (thin solid curve), $2\pi/3$
(dashed curve) and $\pi$ (dash-dotted curve). The latter coincides
with $I(\phi)$ in the corresponding SNS junction ($h_0=0$).}
\label{Fig2}
\end{figure}

\begin{figure}
\includegraphics[width= 7cm]{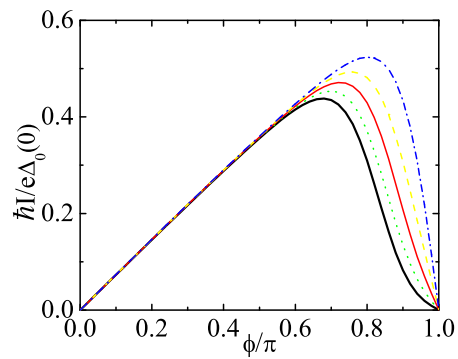}
\caption{(Color online) The current-phase relation $I(\phi)$ in the clean limit for
$T/T_c=0.1$, $h_0/E_{\rm F}=0.1$, $d_1=d_2=500k_{\rm F}^{-1}$,
and five  values of the misorientation angle: $\alpha=0$ (thick solid
curve), $\pi/3$ (dotted curve),  $\pi/2$ (thin solid curve), $2\pi/3$
(dashed curve) and $\pi$ (dash-dotted curve). The latter coincides
with $I(\phi)$ in the corresponding SNS junction ($h_0=0$).}
\label{Fig3}
\end{figure}

\begin{figure}
\includegraphics[width= 6.5cm]{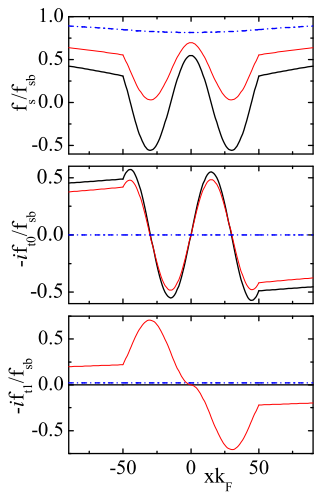}
\caption{(Color online) Spatial dependence of singlet and triplet
pair amplitudes $f_s$, $f_{t0}$ and $f_{t1}$, normalized to the  bulk
singlet pair amplitude $f_{sb}$, for $T/T_c=0.1$, $h_0/E_{\rm
F}=0.1$, $d_1=d_2=50k_{\rm F}^{-1}$, and three  values of the
misorientation angle: $\alpha=0$ (thick solid curve), $\pi/2$ (thin solid curve) and $\pi$ (dash-dotted curve). All amplitudes are
calculated for $\phi=0$ and $\theta=0$.}
\label{Fig4}
\end{figure}

\begin{figure}
\includegraphics[width= 6.5cm]{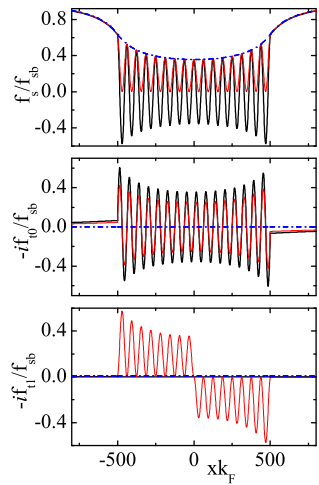}
\caption{(Color online) Spatial dependence of singlet and triplet
pair amplitudes $f_s$, $f_{t0}$ and $f_{t1}$, normalized to the bulk
singlet pair amplitude $f_{sb}$, for $T/T_c=0.1$, $h_0/E_{\rm
F}=0.1$, $d_1=d_2=500k_{\rm F}^{-1}$, and three  values of the
misorientation angle: $\alpha=0$ (thick solid curve), $\pi/2$ (thin solid curve) and $\pi$ (dash-dotted curve).}
\label{Fig5}
\end{figure}

Ballistic regime, $\tau\rightarrow\infty$, is illustrated in Figs.~\ref{Fig2}-\ref{Fig5}. The Josephson current is calculated from  Eq.~(\ref{struja}) for a single transverse channel (the angular averaging $\langle ...\rangle$  reduces to one half of a discrete sum over
$\theta=0$ and $\theta=\pi$, and one-dimensional density of states per spin is $N(0)=1/2\pi \hbar v_F$).
We illustrate the current-phase relations $I(\phi)$ in Figs.~\ref{Fig2} and \ref{Fig3} for five values of the angle between magnetizations in ferromagnetic layers,
$\alpha=0, \pi/3, \pi/2, 2\pi/3,$ and $\pi$.  Previous results obtained in Ref.~\cite{Pajovic} by solving the
Bogoliubov-de Gennes equation are exactly reproduced for this case of transparent interfaces and equal Fermi wave vectors. In particular, we see in Fig.~\ref{Fig2} that the transition
between $0$ and $\pi$ states can be induced by varying the misorientation
angle $\alpha$, including the coexistence of stable and metastable  $0$ and $\pi$ states with dominant second harmonic.\cite{radovic01}  However, nonmonotonic dependence of the Josephson current $I(\phi)$ on $\alpha$ is due to decreasing influence of the exchange potential with increasing misorientation of
magnetizations: $I(\phi)$ for noncollinear magnetizations ($\alpha\neq 0$) is practically equal to the one for homogeneous magnetization ($\alpha=0$) with correspondingly smaller exchange energy. Influence of $\alpha$ on $I(\phi)$ can not be attributed directly to the appearance of odd triplet correlations. When the junction is far enough from the
$0-\pi$ transitions, the critical Josephson current is a monotonic
function of $\alpha$, Fig.~\ref{Fig3}. We emphasize that $I(\phi)$ curves for $\alpha=\pi$ are the same as in the corresponding
SNS junctions ($h_0=0$) since the influence of opposite magnetizations in F layers of a symmetric SFFS junction cancels out in the clean limit and for transparent interfaces.

\begin{figure}
\includegraphics[width= 6.5cm]{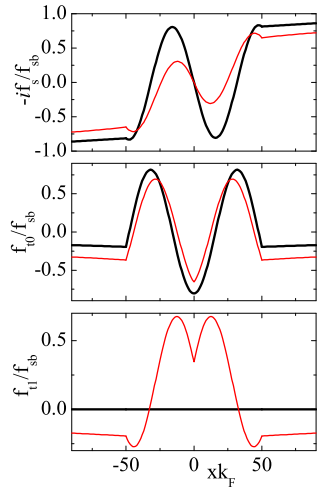}
\caption{(Color online) Spatial dependence of singlet and triplet
pair amplitudes $f_s$, $f_{t0}$ and $f_{t1}$, normalized to the  bulk
singlet pair amplitude $f_{sb}$, for $T/T_c=0.1$, $h_0/E_{\rm
F}=0.1$, $d_1=d_2=50k_{\rm F}^{-1}$, and two  values of the
misorientation angle: $\alpha=0$ (thick solid curve) and $\pi/2$ (thin solid curve). All amplitudes are
calculated for $\phi=\pi$ and $\theta=0$.}
\label{Fig6}
\end{figure}

The spatial variation of the pair amplitudes is shown in Figs.~\ref{Fig4}
and \ref{Fig5} for thin and thick ferromagnetic layers, respectively, for $\theta=0$ and for $\phi=0$.  Note that
$f_s(\theta=\pi)=f_s(\theta=0)$, and
$f_{t1}(\theta=\pi)=f_{t1}(\theta=0)$, while
$f_{t0}(\theta=\pi)=-f_{t0}(\theta=0)$. For $\phi=0$ the singlet amplitude $f_s$ is real, while the triplet
amplitudes $f_{t0}$ and $f_{t1}$ are imaginary (Figs.~\ref{Fig4}
and \ref{Fig5}), and for $\phi=\pi$  the opposite is true (Fig.~\ref{Fig6}). It can be seen that both spin singlet and triplet pair
amplitudes decay in the same oscillatory manner with distance from
the FS interfaces. Similar
behavior of these correlations, i.e.,~the absence of monotonically
decaying long-ranged triplet pair amplitude, which is in a striking
contrast with the dirty-limit case,\cite{bergeret03prltriplet} explains the monotonic
dependence of $I_c$ on $\alpha$ far from the $0-\pi$ transitions. For $\alpha=0$, amplitudes $f_s$ and $f_{t0}$ oscillate with $x$ around zero and $f_{t1}=0$. For $\alpha=\pi/2$ the amplitude $f_{t0}$ oscillates as a function of $x$ as in the previous case, while $f_s$ and $f_{t1}$ oscillate in the same manner above or bellow zero. We find that $f_{t0}$ and $f_{t1}$ oscillate in phase and with the same characteristic length (proportional to $1/h_0$)  as $f_s$, regardless of $\alpha$, when $0<\alpha<\pi$. Phase of  $f_s$ oscillations is shifted by $\pi$.  We also find that triplet correlations penetrate into the superconductors and monotonically decay over  $\xi_S$, the same characteristic length, on which the singlet amplitude saturates. For $\alpha=\pi$, when  the magnetizations are antiparallel no triplet amplitudes exist and $f_s$ monotonically decays away from the FS interfaces as in SNS junctions.

\begin{figure}
\includegraphics[width= 6.5cm]{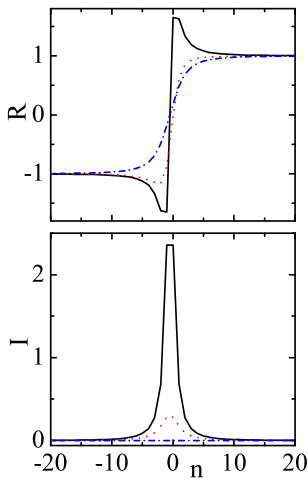}
\caption{(Color online) Real $R$ and imaginary $I$ parts  of  $\langle g_{\uparrow\uparrow}\rangle$ as  functions of $n$ in the Matsubara frequency $\omega_n=\pi k_BT(2n+1)$ for $T/T_c=0.1$, $h_0/E_{\rm
F}=0.1$, $d_1=d_2=50k_{\rm F}^{-1}$, and three  values of the
misorientation angle: $\alpha=0$ (solid curve), $\pi/2$ (dotted curve) and $\pi$ (dash-dotted curve). The curves are calculated in the clean limit, $\tau\rightarrow \infty$, and for  $\phi=0$ and $\theta=0$.}
\label{Fig7}
\end{figure}

\begin{figure}
\includegraphics[width= 6.5cm]{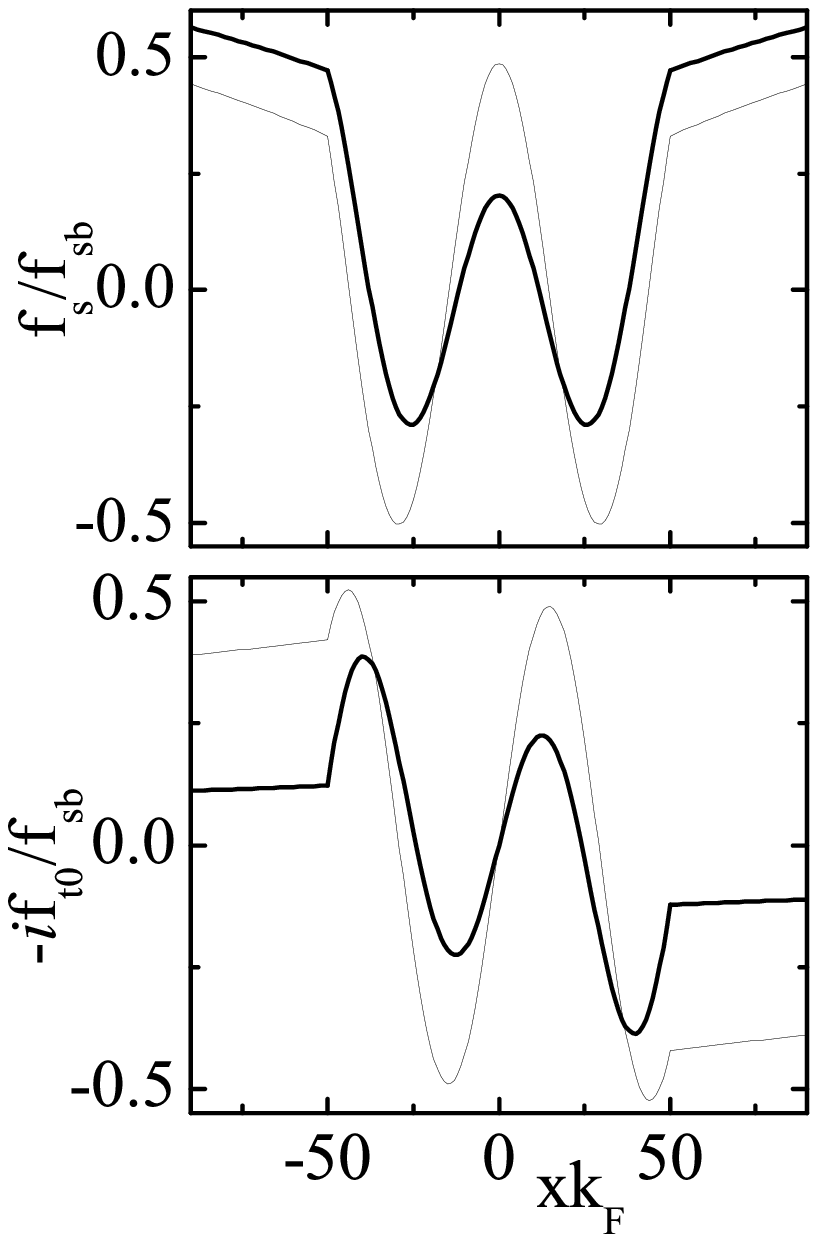}
\caption{Spatial dependence of singlet and triplet
pair amplitudes $f_s$ and $f_{t0}$,  normalized to the bulk singlet
pair amplitude $f_{sb}$, for parallel magnetizations,  $\alpha=0$,
$T/T_c=0.1$, $h_0/E_{\rm F}=0.1$, $d_1=d_2=50k_{\rm F}^{-1}$ and
for  two values of the impurity-scattering rate $\hbar/2\tau
 E_F=0.001$ (thin solid curve) and $0.01$ (thick solid curve).}
\label{Fig8}
\end{figure}

\begin{figure}
\includegraphics[width= 6.5cm]{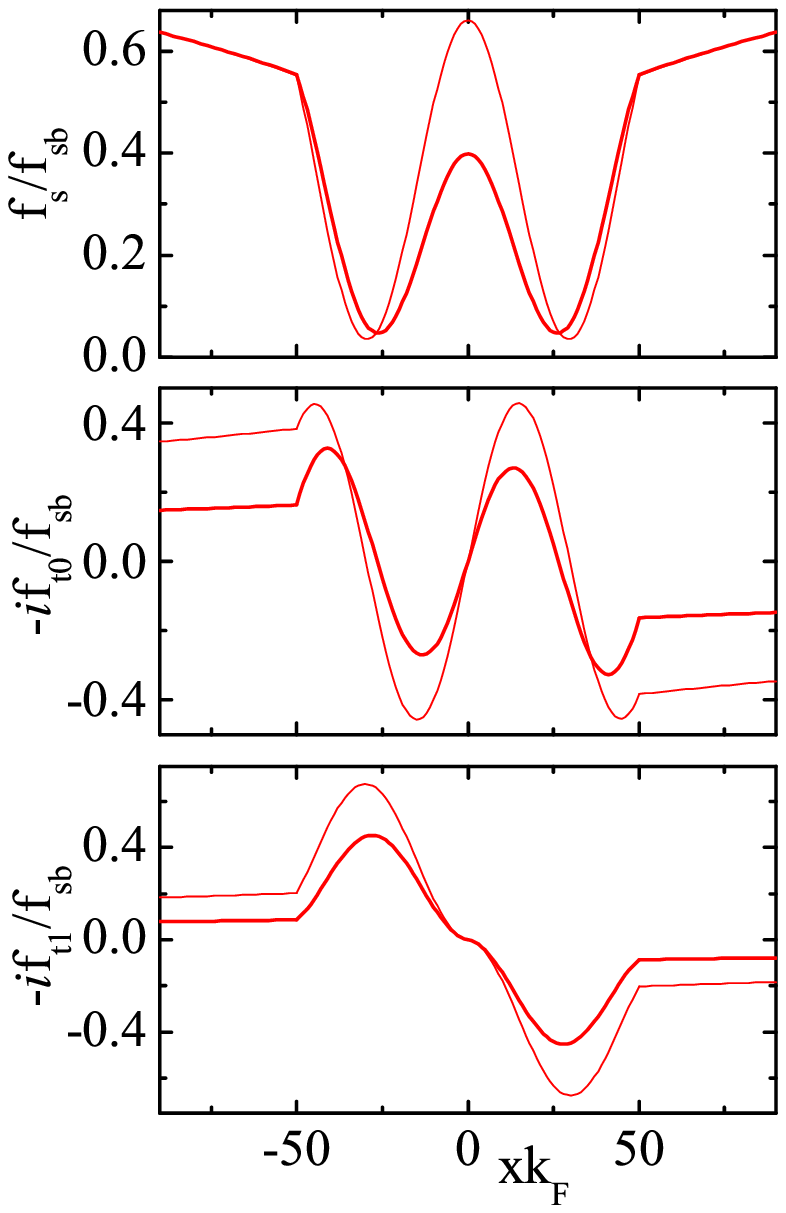}
\caption{(Color online) Spatial dependence of singlet and triplet
pair amplitudes $f_s$, $f_{t0}$ and $f_{t1}$,  normalized to the
bulk singlet pair amplitude $f_{sb}$, for orthogonal magnetizations,
$\alpha=\pi/2$, $T/T_c=0.1$, $h_0/E_{\rm F}=0.1$, $d_1=d_2=50k_{\rm F}^{-1}$ and for  two values of the impurity-scattering rate
$\hbar/2\tau
 E_F=0.001$ (thin solid curve) and $0.01$ (thick solid curve).}
\label{Fig9}
\end{figure}

\begin{figure}
\includegraphics[width= 7cm]{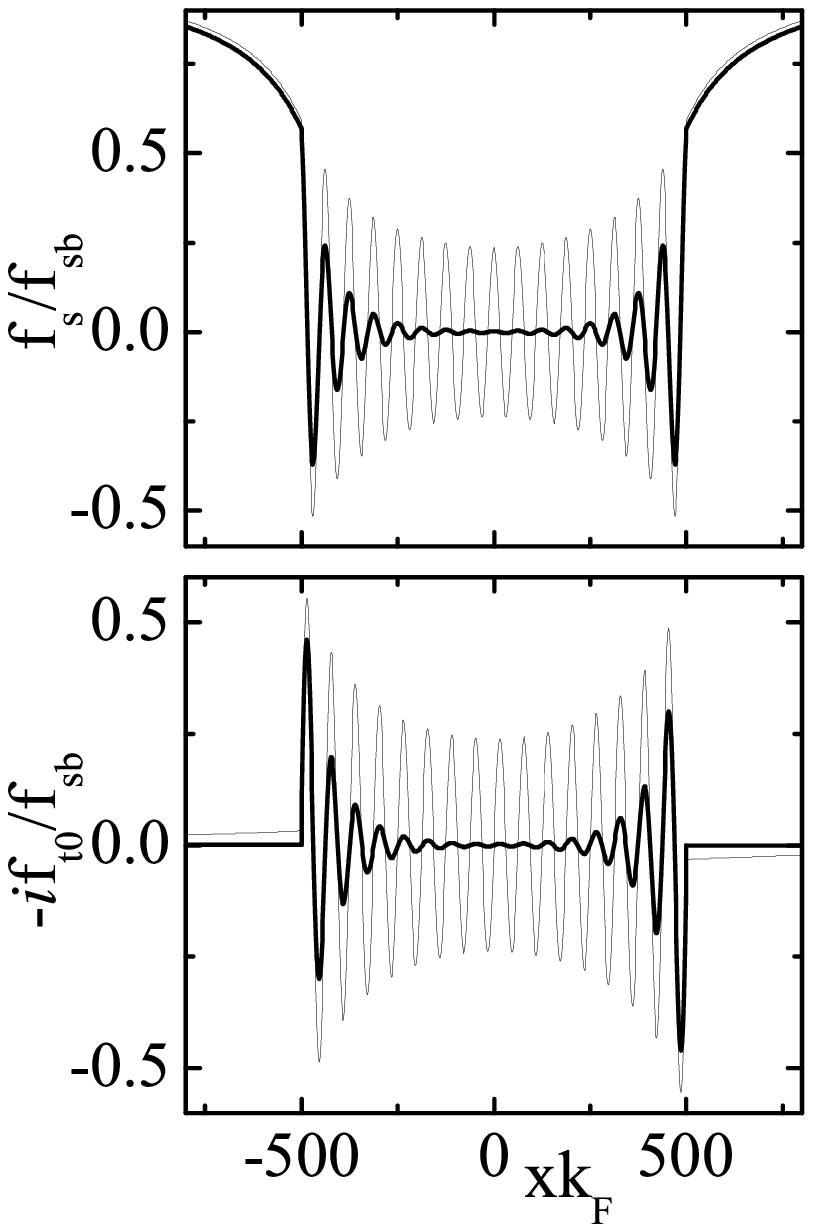}
\caption{Spatial dependence of singlet and triplet
pair amplitudes $f_s$ and $f_{t0}$,  normalized to the bulk singlet
pair amplitude $f_{sb}$, for parallel magnetizations,  $\alpha=0$,
$T/T_c=0.1$, $h_0/E_{\rm F}=0.1$, $d_1=d_2=500k_{\rm F}^{-1}$
and for  two values of the impurity-scattering rate $\hbar/2\tau
 E_F=0.001$ (thin solid curve) and $0.01$ (thick solid curve).}
\label{Fig10}
\end{figure}

\begin{figure}
\includegraphics[width= 6.5cm]{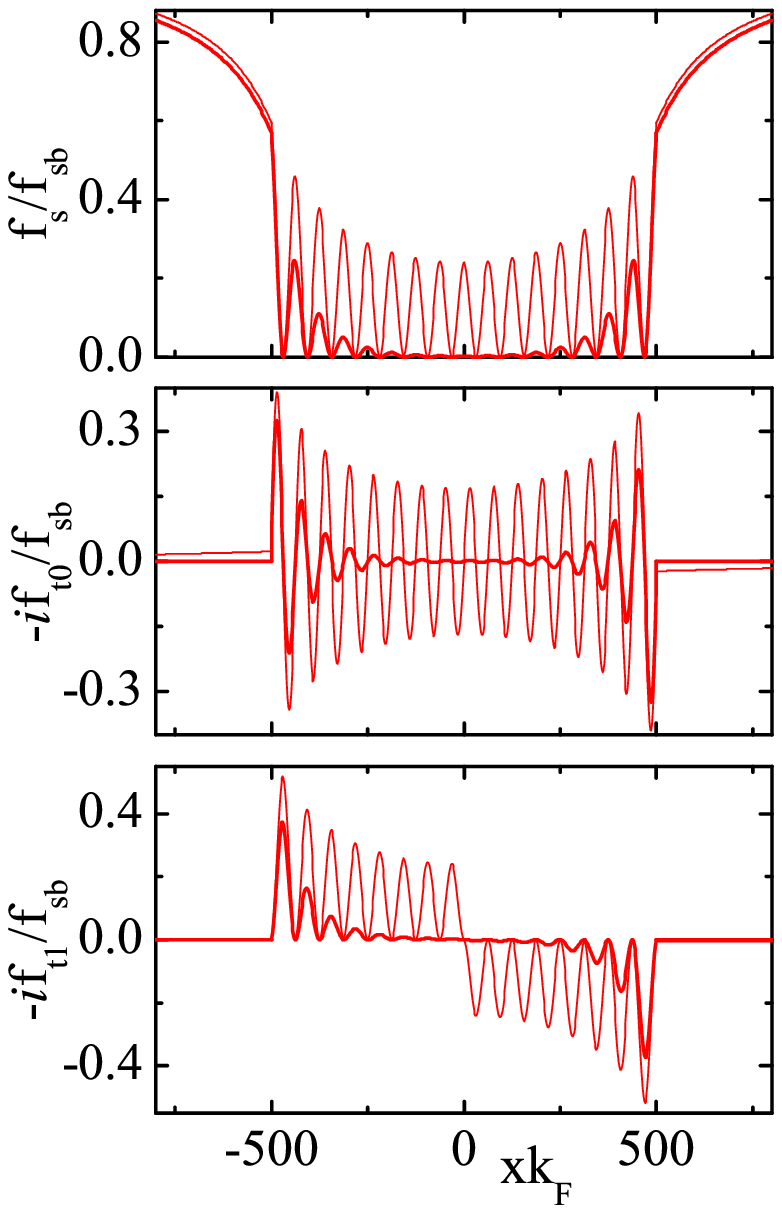}
\caption{(Color online) Spatial dependence of singlet and triplet
pair amplitudes $f_s$, $f_{t0}$ and $f_{t1}$,  normalized to the
bulk singlet pair amplitude $f_{sb}$, for orthogonal magnetizations,
$\alpha=\pi/2$, $T/T_c=0.1$, $h_0/E_{\rm F}=0.1$, $d_1=d_2=500k_{\rm F}^{-1}$ and for  two values of the impurity-scattering rate
$\hbar/2\tau
 E_F=0.001$ (thin solid curve) and $0.01$ (thick solid curve).}
\label{Fig11}
\end{figure}

In order to study the effect of finite $\tau$, we have calculated $\langle g_{\sigma\sigma}\rangle$ in the clean limit ($\tau\rightarrow \infty$) as the largest contribution to the term
$\hbar\langle\check{g}\rangle/2\tau$ in Eq.~(\ref{Eil}).
In Fig.~\ref{Fig7} the typical dependence of real and imaginary parts of $\langle g_{\sigma\sigma}\rangle$ on the Matsubara frequencies is illustrated for thin ferromagnetic layers and three values of $\alpha=0,\pi/2,\pi$. Takeing
$R\approx$ sign$(\omega_n)$ for any $\alpha$, and neglecting contribution of  $I$ which is
nontrivial only in the vicinity of $\omega_n=0$, we calculate $\check{g}$ from the clean limit solutions, Eqs.~(\ref{gF})--(\ref{fF}), replacing only $\kappa,\kappa_{\pm}$  with $\tilde{\kappa},\tilde{\kappa}_{\pm}$, given by Eqs.~(\ref{k11})--(\ref{k13}).

\begin{figure}[h!!!]
\includegraphics[width= 7cm]{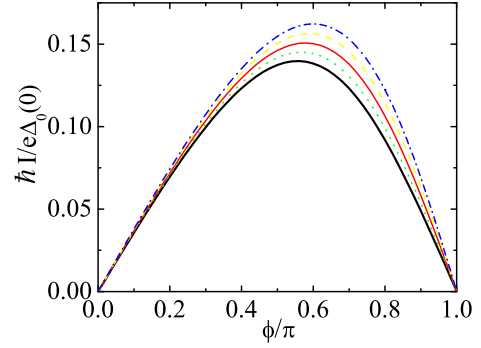}
\caption{(Color online) The current-phase relation $I(\phi)$ for weakly disordered ferromagnets $\hbar/2\tau
 E_F=0.001$, for
$T/T_c=0.1$, $h_0/E_{\rm F}=0.1$, $d_1=d_2=500k_{\rm F}^{-1}$,
and five  values of the misorientation angle: $\alpha=0$ (thick solid
curve), $\pi/3$ (dotted curve),  $\pi/2$ (thin solid curve), $2\pi/3$
(dashed curve) and $\pi$ (dash-dotted curve).}
\label{Fig12}
\end{figure}

\begin{figure}[h!!!]
\includegraphics[width= 7cm]{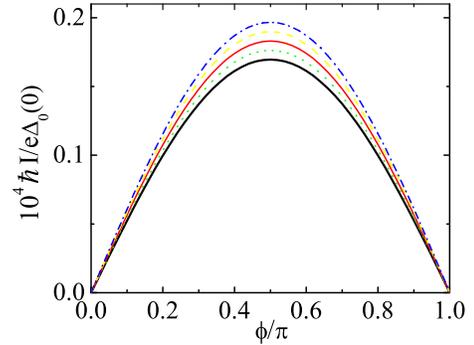}
\caption{(Color online) The current-phase relation $I(\phi)$ for weakly disordered ferromagnets $\hbar/2\tau
 E_F=0.01$, for
$T/T_c=0.1$, $h_0/E_{\rm F}=0.1$, $d_1=d_2=500k_{\rm F}^{-1}$,
and five  values of the misorientation angle: $\alpha=0$ (thick solid
curve), $\pi/3$ (dotted curve),  $\pi/2$ (thin solid curve), $2\pi/3$
(dashed curve) and $\pi$ (dash-dotted curve).}
\label{Fig13}
\end{figure}

 The influence of moderate disorder in ferromagnets on the pair amplitudes
 is illustrated in Figs.~\ref{Fig8}-\ref{Fig11} for the parallel ($\alpha=0$) and the orthogonal
 ($\alpha=\pi/2$) magnetizations,  for two values of the scattering rates, $\hbar/2\tau
 E_F=0.001$ and $0.01$. It can be seen that the only difference from the $\tau\rightarrow \infty$ case is the faster decay with the distance from FS interfaces. However, all amplitudes, singlet and triplet, decay with the same characteristic length. We can not see that the triplets are longer-ranged than the singlets. The influence of scattering on impurities in ferromagnets on the Josephson current is shown in Fig.~\ref{Fig12}, for thick ferromagnetic layers and for $\hbar/2\tau
 E_F=0.001$. Comparison with Fig.~\ref{Fig3}
shows that current-phase relation is almost sinusoidal and the critical current is three times smaller. For ten times larger scattering rate,  $\hbar/2\tau
 E_F=0.01$, numerical calculation gives exactly sinusoidal $I(\phi)$ and four order of magnitudes smaller critical current, Fig.~\ref{Fig13}.

\section{Conclusion}

We have studied the Josephson effect in clean  and moderately disordered SFFS
junctions containing conventional ($s$-wave) superconductors, two
mono-domain ferromagnetic layers with arbitrary angle $\alpha$ between in-plane magnetizations, and fully transparent interfaces. We have calculated the  Josephson current and both singlet and odd in frequency triplet amplitudes in the clean limit and for moderate disorder in ferromagnets by solving the Eilenberger quasiclassical equations analytically  in the step-wise (non self-consistent) approximation for the pair potential. Using quasiclassical approach we limited our considerations to the case of equal Fermi wave vectors, relatively weak ferromagnets and transparent interfaces. For the Josephson current we reproduced previous results for the clean limit~\cite{Pajovic} and have found crossover to the dirty limit case for finite impurity-scattering rate which leads to sinusoidal current-phase relation and considerably smaller value of the critical current. In particular, transitions between $0$ and $\pi$ states are induced by varying the relative orientation of magnetizations, both in clean and diffusive junctions.~\cite{Pajovic,Golubov02,bergeret03prbtriplet} However, in the clean limit this is simply the effect of decreasing influence of the exchange potential with increasing misorientation of magnetizations and far from $0-\pi$ transitions the critical Josephson current monotonically depends on the angle between magnetizations. This can be seen in the calculated spatial variation of the pair amplitudes. We find that for $0<\alpha<\pi$, both spin singlet and triplet pair
amplitudes in F layers decay in the same oscillatory manner with distance from
the FS interfaces. This decay gets faster as the impurity-scattering rate in ferromagnets is increased, but characteristic length of oscillations is unchanged. The computed triplet amplitude $f_{t1}$ has opposite signs in the two magnet regions, penetrates into the superconductors and monotonically decays over the same distance, which is the superconducting coherence length, as the singlet amplitude $f_s$ saturates to the bulk value. In contrast to the dirty limit case,~\cite{reviewbergeret} no substantial impact of odd in frequency spin-triplet superconducting correlations on the Josephson current has been found in the ballistic and  moderately diffusive regimes.

\section{Acknowledgment}

 ZR thanks Gerd Sch\"{o}n, Yasuhiro Asano, and Ivan Bo\v zovi\' c  for useful discussions. The work was supported  by the Serbian Ministry of Science, Project No.~141014.

\end{document}